\documentclass[11pt]{article}
\usepackage{graphicx}
\def \BDM {\begin{displaymath}}
\def \EDM {\end{displaymath}}
\def \BEQ {\begin{equation}}
\def \EEQ {\end{equation}}
\def \BEQA {\begin{eqnarray}}
\def \EEQA {\end{eqnarray}}

\def \BL {\begin{list}}
\def \EL {\end{list}}
\def \BENUM {\begin{enumerate}}
\def \EENUM {\end{enumerate}}
\def \BITEM {\begin{itemize}}
\def \EITEM {\end{itemize}}
\def \BARR {\begin{array}}
\def \EARR {\end{array}}

\begin{document}

%\DeclareGraphicsExtensions{.jpg,.pdf,.mps,.png}

% Title information.
\title{Parallel Programming with MatlabMPI
\thanks{
This work is sponsored by the High Performance Computing Modernization
Office, under Air Force Contract
F19628-00-C-0002.  Opinions, interpretations, conclusions and
recommendations are those of the author and are not necessarily endorsed
by the Department of Defense.
}}

\author{Jeremy Kepner  (kepner@ll.mit.edu)
MIT Lincoln Laboratory, Lexington, MA  02420 \\
}
\maketitle

%%%%%%%%%%%%%%%%%%%%%%%%%%%%%%%%%%%%%%%%%%%%%%%%%%%%%%%%%%%%
\begin{abstract}
%%%%%%%%%%%%%%%%%%%%%%%%%%%%%%%%%%%%%%%%%%%%%%%%%%%%%%%%%%%%

  MatlabMPI is a Matlab implementation of the Message Passing Interface
(MPI) standard and allows any Matlab program to exploit multiple
processors. MatlabMPI currently implements the basic six functions that
are the core of the MPI point-to-point communications standard. The key
technical innovation of MatlabMPI is that it implements the widely used
MPI ``look and feel'' on top of standard Matlab file I/O, resulting in
an extremely compact ($\sim$100 lines) and ``pure'' implementation which runs
anywhere Matlab runs.  The performance has been tested on both shared
and distributed memory parallel computers.  MatlabMPI can match the
bandwidth of C based MPI at large message sizes. A test image filtering
application using MatlabMPI achieved a speedup of $\sim$70 on a parallel
computer.

\end{abstract}

%%%%%%%%%%%%%%%%%%%%%%%%%%%%%%%%%%%%%%%%%%%%%%%%%%%%%%%%%%%%
\section{Introduction}
%%%%%%%%%%%%%%%%%%%%%%%%%%%%%%%%%%%%%%%%%%%%%%%%%%%%%%%%%%%%

  Matlab \cite{Matlab} is the dominant programming language for
implementing numerical computations and is widely used for algorithm
development, simulation, data reduction, testing and system evaluation. 
Many of these computations can benefit from faster execution on a
parallel computer. There have been many previous attempts to provide an
efficient mechanism for running Matlab programs on parallel computers
\cite{MATABP,Morrow98,ParAl,RTExpress,Tseng99,MultiMATLAB,
      ParaMat,Fabozzi99,Matpar,MPITB,Quinn,CMTM}.
These efforts of have faced numerous challenges and none have
received widespread acceptance.

  In the world of parallel computing the Message Passing Interface (MPI)
\cite{MPI} is the de facto standard for implementing programs on multiple
processors. MPI defines C language functions for doing point-to-point
communication in a parallel program.  MPI has proven to be an effective
model for implementing parallel programs and is used by many of the
worlds' most demanding applications (weather modeling, weapons
simulation, aircraft design, and signal processing simulation).

  MatlabMPI consists of a set of Matlab scripts that implements a subset
of MPI and allows any Matlab program to be run on a parallel computer.
The key innovation of MatlabMPI is that it implements the widely used
MPI ``look and feel'' on top of standard Matlab file I/O, resulting in a
``pure'' Matlab implementation that is exceedingly small ($\sim$100 lines
of code). Thus, MatlabMPI will run on any combination of computers that
Matlab supports.

%%%%%%%%%%%%%%%%%%%%%%%%%%%%%%%%%%%%%%%%%%%%%%%%%%%%%%%%%%%%
\section{System Requirements}
%%%%%%%%%%%%%%%%%%%%%%%%%%%%%%%%%%%%%%%%%%%%%%%%%%%%%%%%%%%%

  On shared memory systems, MatlabMPI only requires a single Matlab
license since any user is allowed to launch many Matlab sessions. On a
distributed memory system, MatlabMPI requires one Matlab license per
machine. Because MatlabMPI uses file I/O for communication, there must
also be a directory that is visible to every machine (this is usually
also required in order to install Matlab).  This directory defaults to
the directory that the program is launched from, but can be changed
within the MatlabMPI program.

%%%%%%%%%%%%%%%%%%%%%%%%%%%%%%%%%%%%%%%%%%%%%%%%%%%%%%%%%%%%
\section{Performance Test Results}
%%%%%%%%%%%%%%%%%%%%%%%%%%%%%%%%%%%%%%%%%%%%%%%%%%%%%%%%%%%%

  The vast majority of potential Matlab applications are ``embarrassingly''
parallel and require minimal performance out of MatlabMPI.  These
applications exploit coarse grain parallelism and communicate rarely
(if at all). Never-the-less, measuring performance is useful
for determining which applications are most suitable for MatlabMPI.

\begin{figure}[tbh]
\centerline{\includegraphics[width=4.5in]{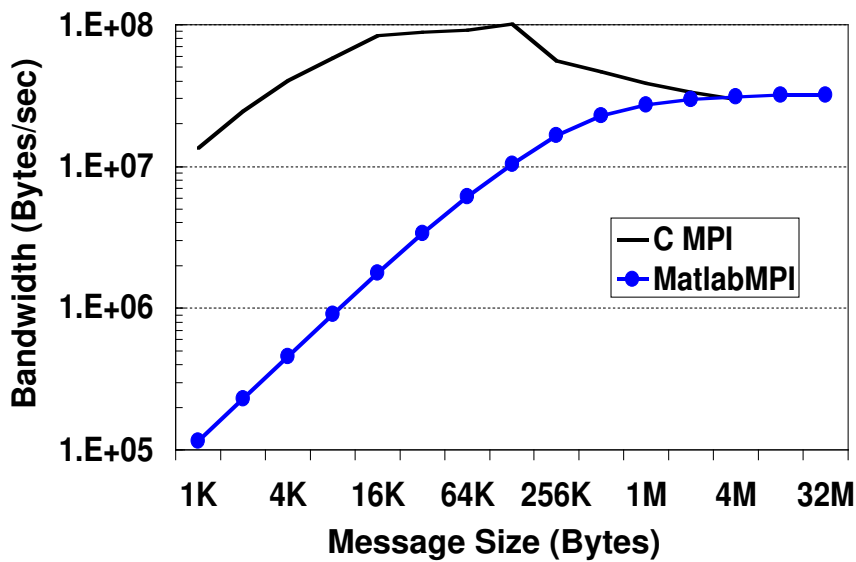}}
\caption{ {\bf Bandwidth.}
  Communication performance as a function message size on the SGI Origin2000.
MatlabMPI equals C MPI performance at large message sizes.
}
\label{fig:bandwidth}
\end{figure}

  MatlabMPI has been run on several Unix platforms.  It has been
benchmarked and compared to the performance of C MPI on the SGI Origin2000. 
These results indicate that for large messages ($\sim$1 MByte)
MatlabMPI is able to match the performance of C MPI (see
Figure~\ref{fig:bandwidth}).  For smaller messages, MatlabMPI is
dominated by its latency ($\sim$35 milliseconds), which is significantly
larger than C MPI.

\begin{figure}[tbh]
\centerline{\includegraphics[width=4.5in]{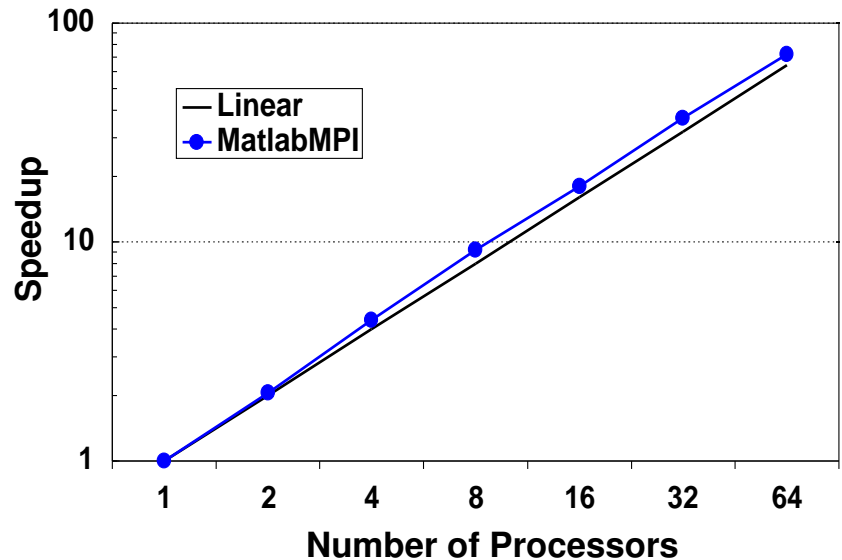}}
\caption{ {\bf Parallel Speedup.}
  Speed increase on the SGI Origin2000 of a parallel image filtering
application as a function of the number of processors. Application shows
``classic'' super-linear performance (due to better cache usage) that
results when a very large memory problem is broken into multiple small
memory problems.
}
\label{fig:speedup}
\end{figure}

  To further test MatlabMPI a test application using MatlabMPI in a
simple image filtering application was written. The application executed
repeated 2D convolutions on a large image (1000 x 128,000 pixels $\sim$
2 GBytes).  This program demonstrates that the MPI standard is valid
within the Matlab environment and allows parallel programs to be written
quickly and easily.  Furthermore, this application achieved excellent
speedups (greater than 64 on 64 processors) and shows the classic
super-linear speedup (due to better cache usage) that comes from
breaking a very large memory problem into many smaller problems (see
Figure~\ref{fig:speedup}).

%%%%%%%%%%%%%%%%%%%%%%%%%%%%%%%%%%%%%%%%%%%%%%%%%%%%%%%%%%%%
\section{Discussion and Future Work}
%%%%%%%%%%%%%%%%%%%%%%%%%%%%%%%%%%%%%%%%%%%%%%%%%%%%%%%%%%%%

   MatlabMPI demonstrates that the standard approach to writing parallel
programs in C and Fortran (i.e. using MPI) is also valid in Matlab.  In
addition, by using Matlab file I/O, it was possible to implement MatlabMPI
entirely within the Matlab environment, making it instantly portable
to all computers that Matlab runs on.  Most potential parallel Matlab
applications are trivially parallel and don't require high performance. 
Never-the-less, MatlabMPI can match C MPI performance on large messages.

  The use of file I/O as a parallel communication mechanism is not new
and is now increasingly feasible with the availability of low cost high
speed disks.  The extreme example of this approach are the now popular
Storage Area Networks (SAN), which combine high speed routers and disks
to provide server solutions. Although using file I/O increases the
latency of messages it normally will not effect the bandwidth. 
Furthermore, the use of file I/O has several additional functional
advantages which make it easy to implement large buffer sizes,
recordable messages, multi-casting, and  one-sided messaging.  Finally,
the MatlabMPI approach is readily applied to any language  (e.g. IDL,
Python, and Perl).

  The simplicity and performance of MatlabMPI makes it a very reasonable
choice for programmers that want to speed up their Matlab code on a
parallel computer.  Further work will aim at enhancing the performance of
MatlabMPI and increase the number MPI functions.

%%%%%%%%%%%%%%%%%%%%%%%%%%%%%%%%%%%%%%%%%%%
% FIGURE CAPTIONS
%%%%%%%%%%%%%%%%%%%%%%%%%%%%%%%%%%%%%%%%%%%

\newpage

%%%%%%%%%%%%%%%%%%%%%%%%%%%%%%%%%%%%%%%%%%%%%%%%%%%%%%%%%%%%

\end{document}